\documentclass[twocolumn,superscriptaddress,amsmath,amssymb,aps,prb]{revtex4-2}

\usepackage{graphicx}
\usepackage{bm}
\usepackage{amsmath}
\usepackage{mathtools}
\usepackage{amssymb}
\usepackage{hyperref}
\usepackage{needspace}
\usepackage[caption=false]{subfig} 

\usepackage{natbib}

\DeclareRobustCommand{\neswarrow}{%
  \mathrel{\text{\ooalign{$\swarrow$\cr$\nearrow$}}}%
}

\begin{document}

\title{Band-minimum degeneracy is not enough: density-of-states control of low-density ferromagnetism}% Force line breaks with \\

\author{Jinghao Wang}
\affiliation{%
 Department of Physics and Astronomy, University of Bologna, Bologna 40126, Italy
}
\affiliation{INFN, Sezione di Bologna, Viale Berti Pichat 6/2, I-40127, Bologna, Italy}

\author{Pierbiagio Pieri}%
\email{pierbiagio.pieri@unibo.it}
\affiliation{%
 Department of Physics and Astronomy, University of Bologna, Bologna 40126, Italy
}
\affiliation{INFN, Sezione di Bologna, Viale Berti Pichat 6/2, I-40127, Bologna, Italy}
%\collaboration{University of Bologna,DIFA}%\noaffiliation

\date{\today}% It is always \today, today,

\begin{abstract}
Recent ultracold-atom experiments have observed ferromagnetic correlations in a geometrically tunable Fermi--Hubbard lattice at intermediate densities. Motivated by subsequent theoretical work that connected the low-density limit of the same model to the M\"uller--Hartmann mechanism for itinerant ferromagnetism, we investigate the stability of ferromagnetism across a broad density range using the $T$-matrix approximation, a controlled approach in the dilute limit. We reproduce the ferromagnetic regime observed experimentally at intermediate densities and systematically compare finite-size and thermodynamic-limit calculations in the dilute regime. We find that the low-density ferromagnetic phase reported for the one-diagonal hopping lattice is strongly suppressed with increasing system size and disappears in the thermodynamic limit, indicating that it originates from finite-size effects. By contrast, low-density ferromagnetism remains robust in the square-lattice Hubbard model with hopping along both diagonal directions. We show that this qualitative difference cannot be explained by the degeneracy of the band minima alone, which occurs in both models. Instead, it is controlled by the singular behavior of the density of states at the band bottom: a weak divergence is insufficient to stabilize ferromagnetism, whereas a much stronger quasi-one-dimensional singularity supports a fully polarized ground state even in the dilute limit. 
Our results demonstrate that band-minimum degeneracy alone is not sufficient for low-density ferromagnetism and that the nature of the band-bottom density-of-states singularity ultimately controls its stability.
\end{abstract}

\maketitle

\section{Introduction}

Itinerant ferromagnetism is one of the most fundamental yet least understood collective phenomena in quantum many-body systems. Despite nearly a century of study, a fully predictive microscopic theory remains elusive \cite{MORIYA19791,arovas2022hubbard,qin2022hubbard}. At the mean-field level, the Stoner criterion predicts ferromagnetism over a broad range of parameters in the Hubbard model \cite{stoner1938collective}. In practice, however, correlation effects substantially restrict the conditions under which a ferromagnetic ground state can emerge \cite{hubbard1963electron,fazekas1999lecture}. This tension is reflected in the fact that rigorous demonstrations of ferromagnetism are limited to a few special situations, including Nagaoka ferromagnetism in the $U\rightarrow\infty$ single-hole limit \cite{nagaoka1966ferromagnetism,tasaki1989extension}, flat-band and nearly flat-band ferromagnetism \cite{mielke1993ferromagnetism,katsura2010ferromagnetism,tasaki2020origin}, multiband mechanisms \cite{tanaka2007metallic,tanaka2016metallic}, and Lieb ferromagnetism on bipartite lattices with unequal sublattice sizes \cite{lieb1989two,gouveia2015magnetic}. Understanding the microscopic ingredients that stabilize ferromagnetism in generic itinerant systems therefore remains a central problem in strongly correlated electron physics.

Particularly intriguing is the possibility of ferromagnetism in the low-density regime. In this limit, the competition between kinetic energy and interactions is especially subtle, and the structure of the single-particle dispersion can play a decisive role. A notable example is the mechanism proposed by M\"uller-Hartmann for the one-dimensional Hubbard model with next-nearest-neighbor hopping \cite{muller1995ferromagnetism}. When a Lifshitz transition splits the band minimum into multiple degenerate minima, the resulting low-energy structure can stabilize a fully polarized ground state at sufficiently low densities. Subsequent numerical studies confirmed the existence of low-density ferromagnetism in this one-dimensional setting \cite{NISHIMOTO,S.DAUL1998,LUHANG}.

Recent developments in ultracold-atom experiments have renewed interest in this question. A Fermi-Hubbard simulator with tunable diagonal hopping has enabled a continuous interpolation between square and triangular lattice geometries and revealed ferromagnetic correlations at intermediate densities \cite{xu2023frustration}. Motivated by these experiments, Li \textit{et al.} investigated the low-density limit of the same lattice model in the strongly interacting regime and reported the emergence of ferromagnetism near a Lifshitz transition where the band minimum becomes degenerate \cite{li2024frustration}. This behavior was interpreted as a two-dimensional realization of the M\"uller-Hartmann mechanism.

This interpretation raises a fundamental question. Is the degeneracy of the band minima by itself sufficient to stabilize low-density ferromagnetism in two dimensions? More generally, what aspects of the low-energy electronic structure determine whether a fully polarized state survives in the thermodynamic limit?

In this work, we address these questions using the $T$-matrix approximation (TMA), a controlled approach in the dilute limit \cite{kanamori1963electron,mattis2012theory}. We study both the optical-lattice model realized experimentally and the conventional square-lattice Hubbard model with next-nearest-neighbor hopping along both diagonal directions. By systematically comparing finite-size calculations with direct thermodynamic-limit evaluations, we distinguish genuine many-body effects from finite-size artifacts.

Our results show that band-minimum degeneracy alone is not sufficient to stabilize low-density ferromagnetism. While we reproduce the ferromagnetic regime observed experimentally at intermediate densities, the low-density ferromagnetic phase previously reported \cite{li2024frustration} for the optical-lattice model is progressively suppressed with increasing system size and disappears in the thermodynamic limit. By contrast, low-density ferromagnetism survives in the conventional square-lattice model. Comparing the two cases reveals that the decisive ingredient is not the degeneracy of the band minima, which occurs in both models, but the singular behavior of the density of states at the band bottom. A sufficiently strong singularity stabilizes a fully polarized state, whereas a weaker divergence does not.

The remainder of this paper is organized as follows. Section~\ref{section two} describes the TMA formalism and the two lattice models. Section~\ref{section three} presents finite-size and thermodynamic-limit phase boundaries. Section~\ref{Divergence} discusses the role of band-bottom density-of-states singularities. Section~\ref{section five} analyzes the evolution of the phase boundary with interaction strength, and Section~\ref{section six} presents our conclusions.
%\needspace{5\baselineskip}
\section{Model and Method}
\label{section two}
\subsection{T-matrix approximation}

Calculations are performed on finite lattices of size $\Omega =L \times L$ with periodic boundary conditions. The electron filling $N=N_\uparrow+N_\downarrow$ is restricted to closed-shell configurations. For fixed $N_\uparrow$ and $N_\downarrow$, the ground-state energy within the $T$-matrix approximation is given by \cite{hlubina1999phase}
\begin{equation}
E = \sum_{\mathbf{k},\sigma} f_{\sigma}(\mathbf{k}) \varepsilon_{\mathbf{k}}
+ \frac{U}{\Omega}\sum_{\mathbf{k},\mathbf{k}'}
\frac{f_{\uparrow}(\mathbf{k})f_{\downarrow}(\mathbf{k}')}{1+U\chi_{pp}(\mathbf{k}+\mathbf{k}',\widetilde{\varepsilon}_{\mathbf{k},\uparrow}+\widetilde{\varepsilon}_{\mathbf{k}',\downarrow})},
\end{equation}
where $f_{\sigma}(\mathbf{k})$ is the Fermi momentum distribution of the noninteracting system,
$\widetilde{\varepsilon}_{\mathbf{k},\sigma}=\varepsilon_{\mathbf{k}}-\mu_\sigma$,
and $\mu_\sigma$ denotes the chemical potential for spin $\sigma$. The particle-particle susceptibility is defined as
\begin{equation}
\chi_{pp}(\mathbf{q},\omega)
=
\frac{1}{\Omega}
\sum_{\mathbf{p}}
\frac{
[1-f_{\uparrow}(\mathbf{p})]
[1-f_{\downarrow}(-\mathbf{p}+\mathbf{q})]
}{
\widetilde{\varepsilon}_{\mathbf{p},\uparrow}
+
\widetilde{\varepsilon}_{-\mathbf{p}+\mathbf{q},\downarrow}
-\omega
}.
\end{equation}

For each filling, the ground-state energy is evaluated for all possible spin polarizations. The paramagnetic state is defined as the configuration with the smallest value of $|N_\uparrow-N_\downarrow|$ that minimizes the total energy. This approach was originally introduced by Kanamori as the ``low density approximation''~\cite{kanamori1963electron}. We refer to it here as the $T$-matrix approximation (TMA). Throughout this work, we compare the fully polarized (Nagaoka) state with the paramagnetic state to determine the magnetic ground state at each filling.

This method becomes highly accurate at low electron densities, because in this regime the dominant effect of electron correlations is the modification of electron trajectories due to two-body collisions, a feature that the TMA fully incorporates \cite{kanamori1963electron,mattis2012theory}. Good agreement with quantum Monte Carlo simulations has been found, with deviations between the two calculations below 1\% \cite{hlubina1997ferromagnetism}. 

To distinguish genuine many body effects from finite size artifacts, we further extend the TMA calculation to the thermodynamic limit. In this case, the momentum summation is replaced by an integration over the Brillouin zone,
\begin{align}
E = &\int_{\mathrm{BZ}} \frac{d^2\mathbf{k}}{(2\pi)^2} \sum_{\sigma} f_{\sigma}(\mathbf{k}) \varepsilon_{\mathbf{k}} \nonumber \\
&+ U \iint_{\mathrm{BZ}} \frac{d^2\mathbf{k}}{(2\pi)^2} \frac{d^2\mathbf{k}'}{(2\pi)^2}
\frac{f_{\uparrow}(\mathbf{k}) f_{\downarrow}(\mathbf{k}')}
{1+U\chi_{pp}\bigl(\mathbf{k}+\mathbf{k}', \widetilde{\varepsilon}_{\mathbf{k},\uparrow} + \widetilde{\varepsilon}_{\mathbf{k}',\downarrow}\bigr)}.
\end{align}

Comparing finite cluster and thermodynamic limit calculations allows us to determine whether the ferromagnetic behavior originates from intrinsic correlation effects or from finite size effects.

\subsection{Optical lattice realization}

We apply the TMA to the model of the recently realized ultracold-atom experiment, where spin-$\frac{1}{2}$ fermions are loaded into an optical lattice with anisotropic diagonal hopping~\cite{xu2023frustration}. In this setup, only one of the two next-nearest-neighbor diagonal hoppings is significant, while the hopping along the other diagonal direction is negligible. The Hamiltonian is given by~\cite{li2024frustration}
\begin{equation}
\hat{H}
=
-t\sum_{\langle ij\rangle,\sigma}
\hat{c}_{i\sigma}^\dagger
\hat{c}_{j\sigma}
-t'
\sum_{\langle\langle ij \neswarrow \rangle\rangle,\sigma}
\hat{c}_{i\sigma}^\dagger
\hat{c}_{j\sigma}
+
U\sum_i
\hat{n}_{i\uparrow}
\hat{n}_{i\downarrow},
\end{equation}
where $\sigma=\uparrow,\downarrow$ labels the two spin components. The nearest-neighbor hopping amplitude is denoted by $t$, while $t'$ represents the next-nearest-neighbor hopping along one-diagonal direction. By continuously tuning $t'/t$ from $0$ to $1$, the lattice geometry interpolates between the square ($t'=0$) and triangular lattices ($t'=t$). The on-site repulsive interaction is characterized by $U$, with the experiment performed near $U/t\approx 9$. The corresponding single particle dispersion is
\begin{equation}
\label{oned}
\varepsilon_{\mathbf{k}}
=
-2t(\cos k_x+\cos k_y)
-2t'\cos(k_x+k_y).
\end{equation}
Throughout this work, we set $t=1$ as the unit of energy.

The optical lattice experiment explored the filling range $0.5<n<1.5$ and reported ferromagnetic correlations for $t'/t>0.5$ in the electron-doped regime $1.25<n<1.5$. Owing to the particle-hole symmetry of the one-diagonal hopping model, the electron-doped regime can be mapped onto the hole-doped regime through the transformation $t\rightarrow -t'$, $n\rightarrow 1-n$. In this work, we therefore focus on the hole-doped side $n<1$ and use the dispersion relation given in Eq.~(\ref{oned}) with the corresponding transformed hopping parameters.

\subsection{One-diagonal and two-diagonal hopping models}

For comparison, we also study the conventional square-lattice Hubbard model with both nearest-neighbor and next-nearest-neighbor hoppings along the two diagonal directions. Hereafter, we refer to this as the \emph{two-diagonal hopping model}. The Hamiltonian is given by
\begin{equation} \hat{H}=-t\sum_{\langle ij\rangle,\sigma} \hat{c}_{i\sigma}^\dagger \hat{c}_{j\sigma} -t' \sum_{\langle\langle ij \rangle\rangle,\sigma} \hat{c}_{i\sigma}^\dagger \hat{c}_{j\sigma} + U\sum_i \hat{n}_{i\uparrow} \hat{n}_{i\downarrow}, \end{equation}
and the corresponding dispersion relation is
\begin{equation}
\label{twod}
\varepsilon_{\mathbf{k}}= -2t(\cos k_x+\cos k_y) -4t'\cos k_x\cos k_y.
\end{equation}

Although the two models differ in their hopping geometry, both undergo a Lifshitz transition near $t'/t=-0.5$ and develop degenerate band minima. 

Specifically, for the one-diagonal hopping model, rewwriting the dispersion relation [Eq.~(\ref{oned})] 
in terms of $k_ {\pm} = (k_x \pm k_y)/2$ yields
\begin{equation}
\varepsilon_{\mathbf{k}} = -4t\cos k_+\cos k_- - 2t'\cos 2k_+.
\end{equation}
The minimum lies along the $k_- = 0$ line. While the band exhibits a single minimum  at $k_+ = 0$ for $t'/t > -0.5$, two degenerate minima occur at nonzero $k_+$ when $t'/t < -0.5$.
Thus, a Lifshitz transition\cite{lifshitz1960anomalies} occurs exactly at $t'/t = -0.5$.

A similar analysis of the two-diagonal hopping dispersion [Eq.~(\ref{twod})]  reveals that it undergoes a Lifshitz transition at the same critical ratio. For $t'/t>-0.5$ the band minimum is located at $(0,0)$, while for $t'/t<-0.5$ it splits into four degenerate minima located at $(0,\pm\pi)$ and $(\pm\pi,0)$.

The comparison between the two models therefore provides an ideal setting to investigate whether band-minimum degeneracy alone is sufficient to stabilize low-density ferromagnetism. 
In the following sections, we compare their magnetic phase diagrams from finite clusters to the thermodynamic limit and analyze the microscopic origin of their markedly different behavior in the dilute regime.

\section{Phase diagrams and finite-size effects}
\label{section three}
\begin{figure}[htbp]
    \centering
    \includegraphics[width=1\linewidth]{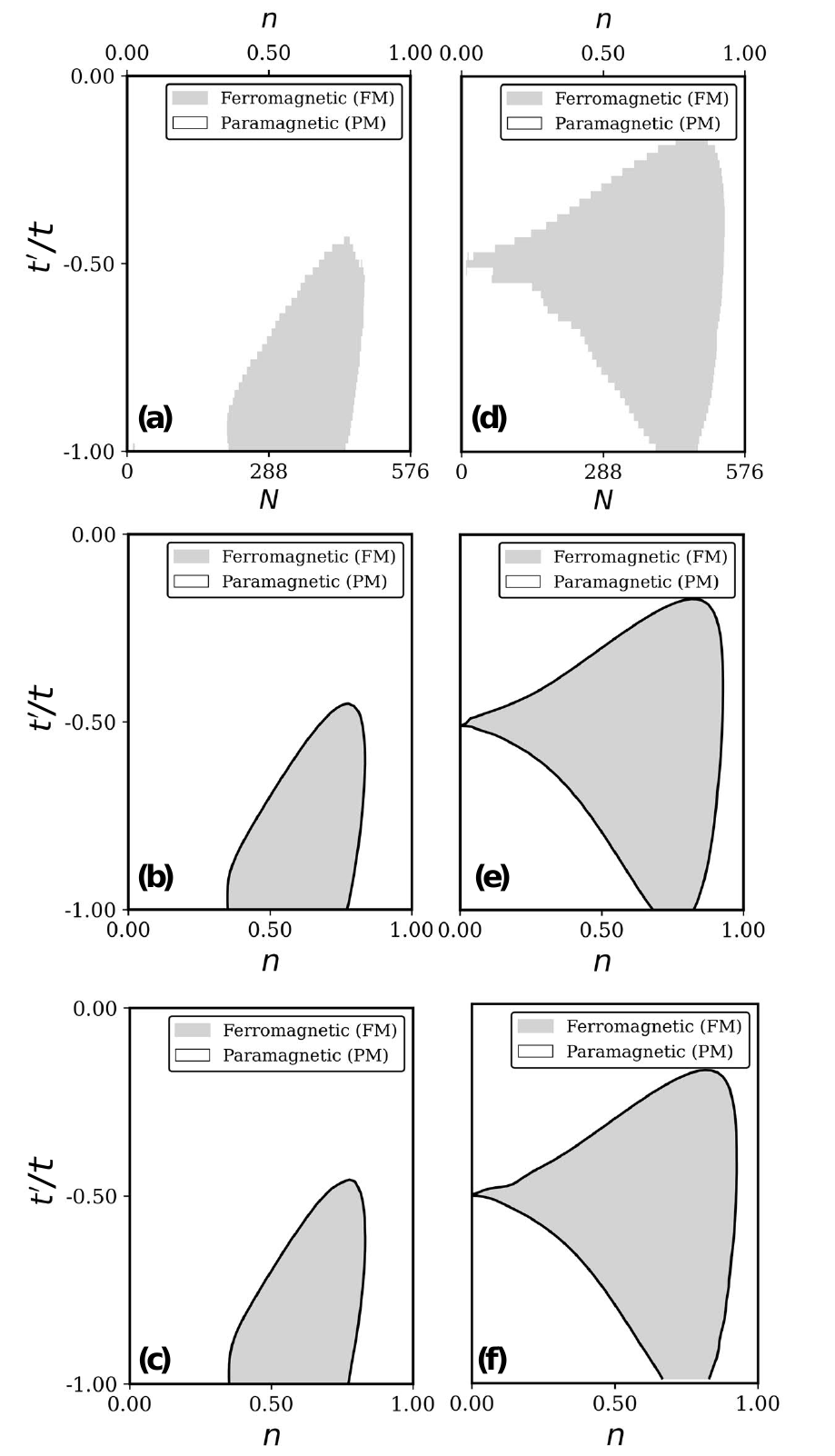}
    \caption{Phase diagrams at $U=9t$. The grey region denotes the stability region of the fully polarized ferromagnetic state obtained within the T-matrix approximation (TMA). one-diagonal hopping lattice: (a) for $L=24$, where the horizontal axis represents the electron number $N$ and $n$ is the filling factor. (b) for $L=100$ and (c) for the thermodynamic limit. two-diagonal hopping lattice: (d) for $L=24$, (e) for $L=100$, and (f) for the thermodynamic limit.}
    \label{fig:fig1} 
\end{figure}
We now compare the magnetic phase diagrams of the one-diagonal and two-diagonal hopping models. Both systems undergo a Lifshitz transition near $t'/t=-0.5$, where the band minimum becomes degenerate. If band-minimum degeneracy were sufficient to stabilize low-density ferromagnetism, one would expect the two models to display similar behavior in the dilute limit. As we show below, this is not the case. We first present the phase diagrams for the same values of the Hubbard repulsion $U=9t$ used in the experiment \cite{xu2023frustration} and then move to the case $U\to+\infty$ considered in the theoretical work~\cite{li2024frustration}.

\subsection{$U=9t$}
Figure~\ref{fig:fig1} summarizes the phase diagrams obtained within the T-matrix approximation at $U=9t$. Panels (a)--(c) correspond to the one-diagonal hopping model, while panels (d)--(f) show the corresponding results for the two-diagonal hopping model. In each case, we compare a small finite cluster, a large finite cluster, and the thermodynamic limit. The grey region denotes the stability region of the fully polarized ferromagnetic state.

For the one-diagonal hopping model, our results reproduce the ferromagnetic regime observed experimentally at intermediate densities. Ferromagnetism appears for sufficiently large negative values of $t'/t$ over a broad density window extending roughly from $n\simeq0.4$ to $n\simeq0.8$. Since the experiment did not explore densities below $n=0.5$, our calculations suggest that ferromagnetic correlations may persist over a somewhat broader density range than currently established.

For the two-diagonal hopping model, our results are consistent with the phase boundary reported in Ref.~\cite{hlubina1999phase}. 
In particular, both our calculations and the results of Ref.~\cite{hlubina1999phase} agree with the low density analysis of Ref.~\cite{pieri1996low}, which showed that in the limit \(n \rightarrow 0\), a fully polarized ferromagnetic ground state can occur only within the range \(-0.2 > t'/t > -0.65\).

The most striking difference between the two models emerges at low density. Although both systems develop degenerate band minima near $t'/t=-0.5$, only the two-diagonal hopping model exhibits a ferromagnetic phase that extends continuously toward $n\rightarrow0$. In the one-diagonal hopping model, by contrast, the ferromagnetic region remains separated from the dilute limit. Thus, the two lattice geometries display qualitatively different behavior despite their similar low-energy band topology.

Finite-size effects are most visible in the small clusters [Figs.~\ref{fig:fig1}(a) and \ref{fig:fig1}(d)], where the phase boundaries display irregular fluctuations associated with the discrete momentum grid. These fluctuations are strongly reduced as the system size increases. The $L=100$ calculations [Figs.~\ref{fig:fig1}(b) and \ref{fig:fig1}(e)] closely reproduce the direct thermodynamic-limit results [Figs.~\ref{fig:fig1}(c) and \ref{fig:fig1}(f)], indicating that this lattice size is sufficient to capture the bulk behavior of both models.

\begin{figure}[t]
    \centering
    \includegraphics[width=1\linewidth]{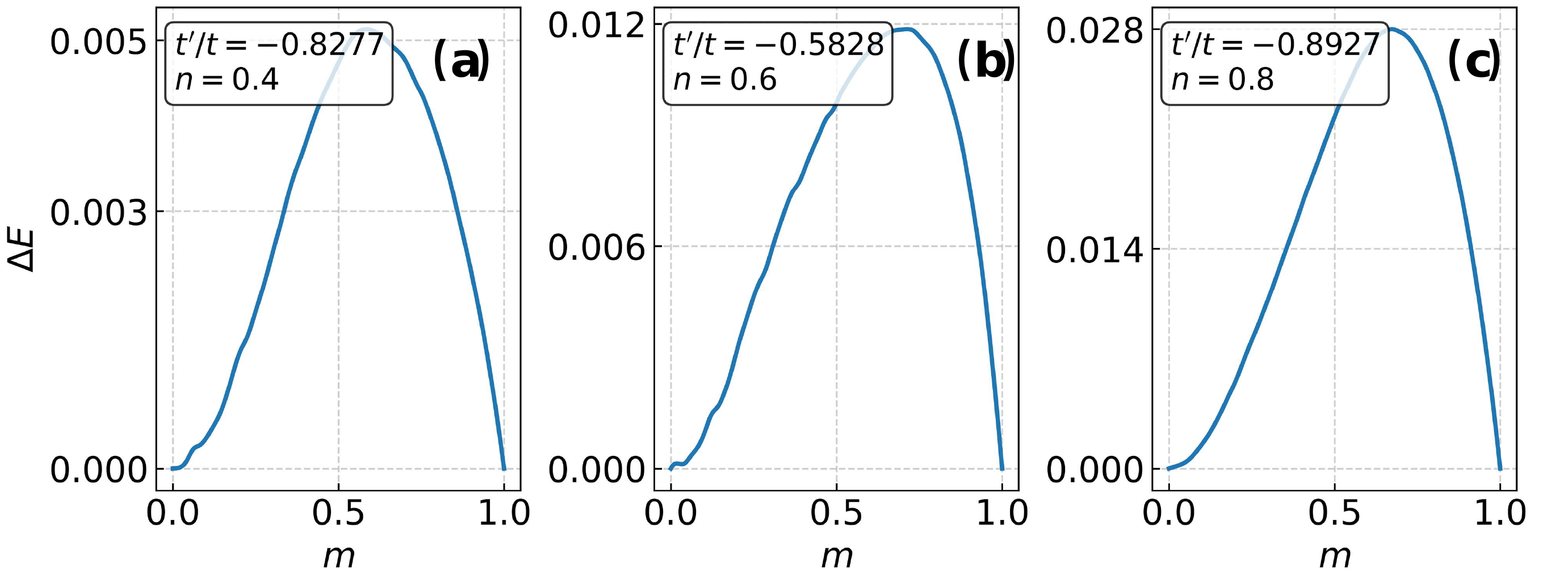}
    \caption{Magnetization $(n_{\uparrow}-n_{\downarrow})/n$ as a function of energy difference $\Delta E$ at different points near the phase boundary for $U=9t$, $L=100$: (a) $t'/t=-0.8277$, $n=0.4$; (b) $t'/t=-0.5828$, $n=0.6$; (c) $t'/t=-0.8927$, $n=0.8$.}
    \label{fig:magnet}
\end{figure}

The phase diagrams shown in Fig.~\ref{fig:fig1} were obtained by comparing the energies of the paramagnetic and fully polarized states. Previous work on the two-diagonal hopping model showed that partially polarized states and phase separation may occur close to the phase boundary without significantly affecting its macroscopic location \cite{hlubina1999phase}. We therefore examine whether the same conclusion applies to the one-diagonal hopping model.

Figure~\ref{fig:magnet} shows the energy difference $\Delta E$ relative to the paramagnetic state as a function of magnetization for representative points near the phase boundary. We find that the transition is governed by a direct level crossing between a paramagnetic state and a fully polarized state. The energy landscape does not develop a minimum at intermediate magnetization, indicating that partially polarized states play only a minor role in determining the macroscopic phase boundary. Consequently, restricting the analysis to the paramagnetic and fully polarized states provides an accurate description of the ferromagnetic region.

\subsection{$U\to+\infty$}
A recent numerical study employed density-matrix renormalization group (DMRG) 
calculations on finite strip geometries to investigate the strongly interacting 
limit ($U\to +\infty$) of the one-diagonal hopping model \cite{li2024frustration}. That 
work reported the emergence of low density ferromagnetism near $t'/t \approx -0.5$, 
which was attributed to the M\"uller-Hartmann mechanism.

M\"uller-Hartmann mechanism is a proposed mechanism for itinerant ferromagnetism in the low-density regime.\cite{muller1995ferromagnetism} This  mechanism considered a one-dimensional Hubbard model with both nearest-neighbor hopping $t$ and next-nearest-neighbor hopping $t'$,
\begin{align} 
\hat{H}={-}\sum_{i \sigma} (t\,\hat{c}_{i,\sigma}^\dagger \hat{c}_{i+1,\sigma} {+} t'\,\hat{c}_{i,\sigma}^\dagger \hat{c}_{i+2,\sigma} {+} \text{H.c.}) + U \sum_i \hat{n}_{i\uparrow}\hat{n}_{i\downarrow}. 
\end{align}

For $t'/t<-1/4$, the band minimum splits into two degenerate minima. M\"uller-Hartmann argued with the construction of a mapping to an effective model that this doubly degenerate band-bottom structure stabilizes a fully polarized ferromagnetic ground state at sufficiently low electron densities. The occurrence of low-density ferromagnetism in this one-dimensional model was later confirmed numerically in Refs. ~\cite{pieri1996low,NISHIMOTO,S.DAUL1998,LUHANG}

Ref.~\cite{li2024frustration} proposed that the two-dimensional one-diagonal hopping model 
harbors a direct analog of this mechanism, due to the occurrence of degenerate minima when $t'/t \le -0.5$. 

\begin{figure}[t]
    \centering
    \includegraphics[width=1\linewidth]{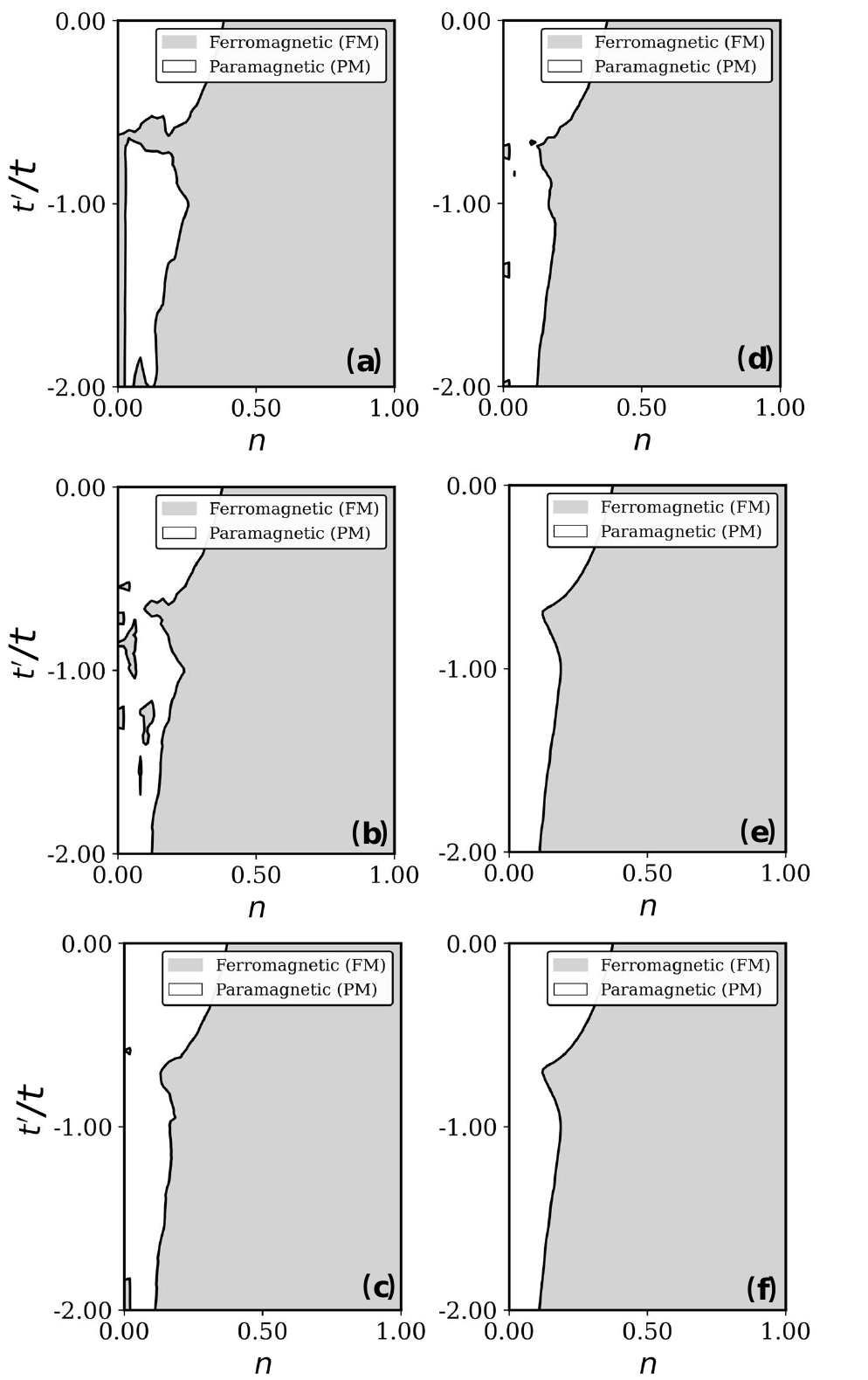}
    \caption{Phase diagrams for the one-diagonal hopping model at infinite interaction strength ($U\to +\infty$) across different cluster sizes: (a) $6 \times 20$, (b) $10 \times 20$, and (c) $16 \times 20$ finite clusters; and square clusters of size (d) $L=20$, (e) $L=48$, and (f) $L=56$.}
    \label{fig:fig2}
\end{figure}

To rigorously test the robustness of this $U\to +\infty$ ferromagnetic phase, we perform 
TMA calculations matching the geometries and interactions of the DMRG study. As shown 
in Fig.~\ref{fig:fig2}(a), when simulating small clusters (e.g., $6 \times 20$) comparable 
to the strips used in DMRG, we indeed reproduce the low density ferromagnetic region. 
However, as we systematically increase the system size [Figs.~\ref{fig:fig2}(b)–(d)], 
this region progressively shrinks. For sufficiently large clusters [Figs.~\ref{fig:fig2}(e) 
and \ref{fig:fig2}(f)], it vanishes entirely. Consistent with our findings at $U=9t$, 
this systematic scaling behavior strongly indicates that the low density ferromagnetism 
in the one-diagonal hopping model is a finite size artifact rather than a robust 
thermodynamic phase.

\begin{figure}[t]
    \centering
    \includegraphics[width=1\linewidth]{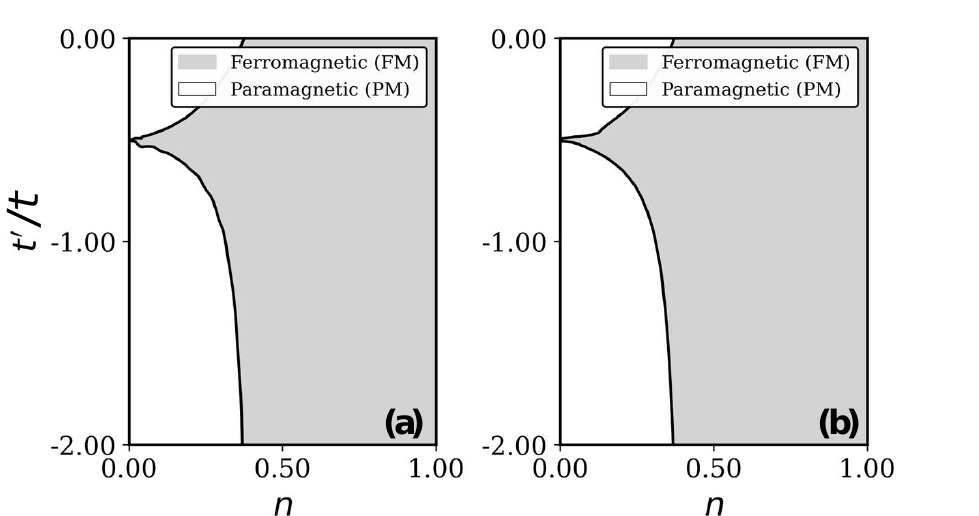}
    \caption{Phase diagrams for the two-diagonal hopping model at infinite interaction strength ($U\to +\infty$): (a) large finite cluster with $L=48$, and (b) in the thermodynamic limit.}
    \label{fig:fig3}
\end{figure}

By contrast, when we apply the same $U\to +\infty$ analysis to the two-diagonal hopping 
model, the results are qualitatively different. As shown in Fig.~\ref{fig:fig3}, both 
large-cluster calculations ($L=48$) and direct thermodynamic limit evaluations reveal 
a persistent low density ferromagnetic region near $t'/t \approx -0.5$. The stability 
of this phase against system-size variation confirms that the ferromagnetism in the 
two-diagonal hopping model is an intrinsic many body effect.

The results of the present section reveal a central puzzle. Both lattice geometries undergo a Lifshitz transition and develop degenerate band minima near $t'/t=-0.5$, yet only one of them supports low-density ferromagnetism in the thermodynamic limit. Understanding the origin of this qualitative difference requires a closer examination of the low-energy structure of the two models, which we undertake in the next section.

\section{Origin of the difference between the two models}
\label{Divergence}

The results of Sec.~\ref{section three} show that band-minimum degeneracy alone does not determine the stability of low-density ferromagnetism. Both lattice geometries undergo a Lifshitz transition near $t'/t=-0.5$ and develop degenerate band minima, yet only the two-diagonal hopping model supports low-density ferromagnetism in the thermodynamic limit. We now investigate the microscopic origin of this difference.

To elucidate the microscopic origin of this discrepancy, we compute the density of states (DOS) near the band bottom $t'/t = -0.5$. The DOS is formally defined as
\begin{equation}
D(\varepsilon) = \int_{\mathrm{BZ}} \frac{d^2\mathbf{k}}{(2\pi)^2} \delta(\varepsilon - \varepsilon_{\mathbf{k}}).
\end{equation}
As the energy $\varepsilon$ approaches the band minimum $E_{\min}$, both models exhibit a singularity, but the nature of the divergence differs fundamentally. Let $\Delta E = \varepsilon - E_{\min}$ denote the energy measured from the band bottom. 

For the one-diagonal hopping model, integrating Eq.~(\ref{oned}) reveals that the DOS diverges as a weak power law:
\begin{equation}
D(\Delta E) = \frac{[\Gamma(1/4)]^2}{2^{11/4} \pi^{5/2}} (\Delta E)^{-1/4},
\end{equation}
for $\Delta E \to 0$.
In stark contrast, evaluating the DOS for the two-diagonal hopping model using Eq.~(\ref{twod}) yields a much stronger singularity:
\begin{equation}
D(\Delta E) \approx \frac{1}{2\sqrt{2}\pi^2 \sqrt{\Delta E}} \ln\left( \frac{128}{\Delta E} \right).
\end{equation}
for $\Delta E \to 0.$
This divergence is dominated by the $(\Delta E)^{-1/2}$ factor, which is strongly reminiscent of the singular DOS typically found in one-dimensional systems.

In Ref.~\cite{pieri1996low} it was shown that low-density ferromagnetism is ruled out in dimension $d \ge 3$, while in $d=2$ it is allowed under condition of a large enough DOS at the bottom of the band, and in $d=1$ low-density ferromagnetism is always possible. 
We attribute the occurrence of ferromagnetism in the two-diagonal hopping case with $t'/t=-0.5$ to an effective reduction of dimensionality from $d=2$ to $d=1$.

The comparison between the two models therefore provides a direct counterexample to the idea that band-minimum degeneracy alone guarantees low-density ferromagnetism. Although both systems undergo the same Lifshitz transition and develop degenerate band minima, they exhibit fundamentally different density-of-states singularities. The relatively weak divergence of the one-diagonal hopping model is insufficient to stabilize a fully polarized state in the thermodynamic limit. By contrast, the much stronger quasi-one-dimensional singularity of the two-diagonal hopping model effectively enhances the low-energy phase space and supports robust low-density ferromagnetism. In this sense, the decisive ingredient is not the degeneracy of the band minima itself, but the nature of the density-of-states singularity that accompanies it.

\section{Evolution of the ferromagnetic phase boundary}
\label{section five}

\begin{figure}[t]
    \centering
    \includegraphics[width=1\linewidth]{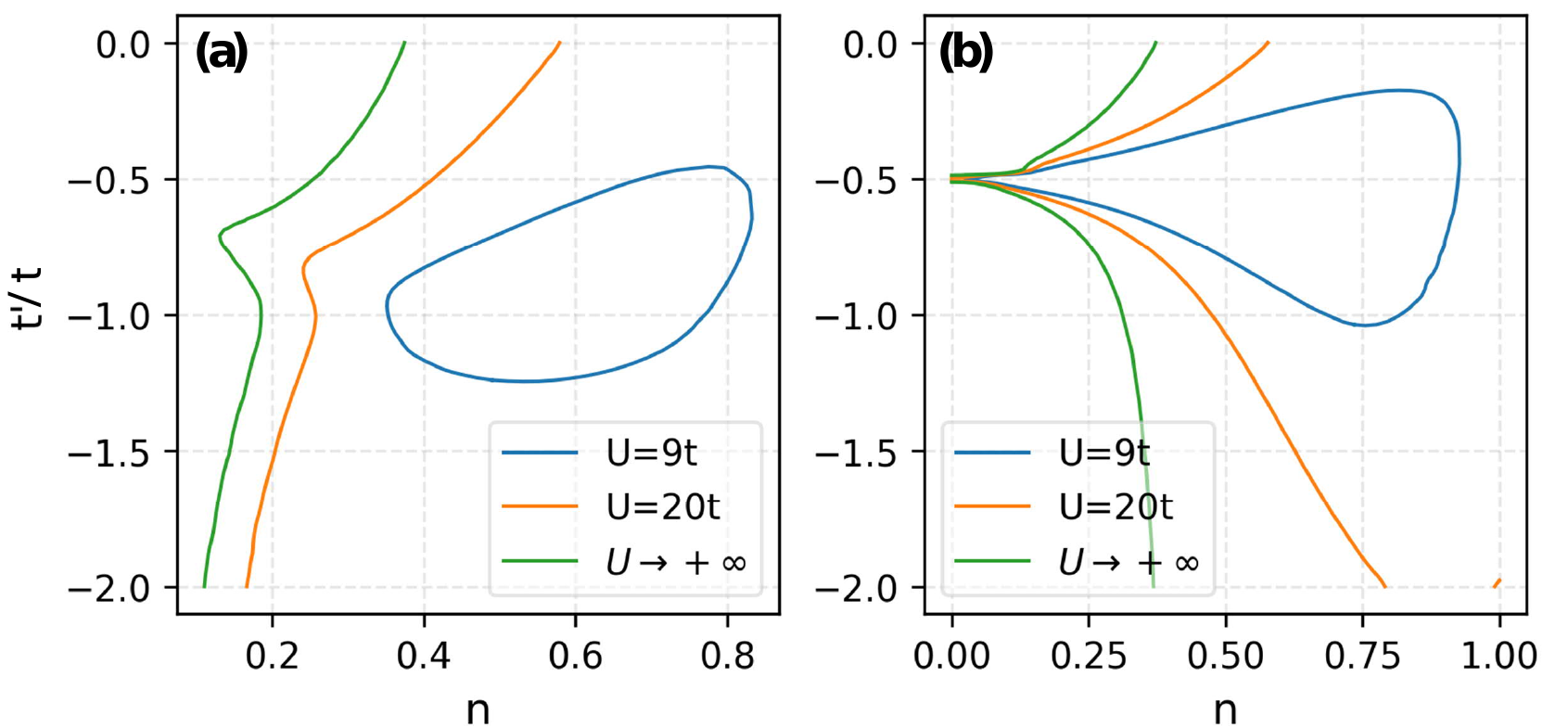}
    \caption{Evolution of the ferromagnetic phase boundary as a function of electron filling $n$ and next-nearest-neighbor hopping $t'/t$ for three interaction strengths $U = 9t$, $20t$, and $U\to +\infty$, obtained from thermodynamic limit TMA calculations. (a) one-diagonal hopping model. (b) two-diagonal hopping model.}
    \label{fig:fig4}
\end{figure}

In the previous section we identified the nature of the band-bottom density-of-states singularity as the key factor controlling the stability of low-density ferromagnetism. We now examine how this mechanism evolves with interaction strength by tracking the ferromagnetic phase boundaries for $U=9t$, $20t$, and $U\rightarrow+\infty$ in the thermodynamic limit.

In Fig. \ref{fig:fig4} we track the evoloution of the phase buonadaries for varying interaction strengths ($U = 9t$, $20t$, and $U\to +\infty$) for both models in the thermodynamics limit. The evolution of the phase boundaries reveals a qualitative distinction between the two lattice geometries. In the one-diagonal hopping model [Fig.~\ref{fig:fig4}(a)], the ferromagnetic region expands monotonically with increasing interaction strength. However, even in the limit $U\rightarrow+\infty$, the phase boundary terminates at a finite density and never reaches the dilute limit.

By contrast, the two-diagonal hopping model [Fig.~\ref{fig:fig4}(b)] exhibits a ferromagnetic region that extends to arbitrarily low density already at moderate interaction strengths. Remarkably, the low-density boundary near $t'/t\simeq -0.5$ remains almost unchanged as $U$ increases from $9t$ to $+\infty$.

This behavior provides further evidence that band-minimum degeneracy alone does not control the stability of low-density ferromagnetism. If degeneracy were the decisive ingredient, one would expect the low-density boundary to evolve similarly in the two models. Instead, the dramatically different response to increasing interaction strength reflects the different density-of-states singularities discussed in Sec.~\ref{Divergence}.

In the two-diagonal hopping model, the strong quasi-one-dimensional singularity already produces a large low-energy phase space at moderate interaction strengths, making the low-density ferromagnetic region nearly insensitive to further increases of $U$. In the one-diagonal hopping model, where the singularity is much weaker, no such saturation occurs. The interaction strength enlarges the ferromagnetic region, but cannot stabilize ferromagnetism in the dilute limit even when $U\rightarrow+\infty$.

\section{Conclusion}
\label{section six}

We have investigated whether band-minimum degeneracy is sufficient to stabilize low-density ferromagnetism in two-dimensional Hubbard models. To address this question, we compared two lattice geometries that both undergo a Lifshitz transition near $t'/t=-0.5$ and develop degenerate band minima, while exhibiting markedly different magnetic behavior in the dilute limit.

Using the $T$-matrix approximation, we first reproduced the ferromagnetic regime observed experimentally at intermediate densities in the one-diagonal hopping optical-lattice model. We then systematically compared finite-size and thermodynamic-limit calculations and found that the low-density ferromagnetic phase previously reported for this model is progressively suppressed with increasing system size and disappears in the thermodynamic limit. By contrast, the two-diagonal hopping model supports a robust low-density ferromagnetic phase that persists from moderate interactions ($U=9t$) to the strongly correlated limit ($U\rightarrow+\infty$).

The comparison between these two models reveals that band-minimum degeneracy alone does not determine the stability of low-density ferromagnetism. Although both systems develop degenerate band minima at the same Lifshitz transition, they exhibit fundamentally different density-of-states singularities at the band bottom. The relatively weak divergence of the one-diagonal hopping model, $D(\Delta E)\propto (\Delta E)^{-1/4}$, is insufficient to stabilize a fully polarized state in the thermodynamic limit. By contrast, the much stronger quasi-one-dimensional singularity of the two-diagonal hopping model, $D(\Delta E)\propto (\Delta E)^{-1/2}\ln(1/\Delta E)$, provides a sufficiently large low-energy phase space to support robust low-density ferromagnetism.

From a broader perspective, our results suggest that the key ingredient controlling low-density ferromagnetism is not the existence of degenerate band minima itself, but the nature of the density-of-states singularity associated with them. In particular, the strong singularity of the two-diagonal hopping model can be viewed as an effective reduction of dimensionality from $d=2$ toward $d=1$, where low-density ferromagnetism is known to be particularly favorable. The comparison presented here therefore provides a direct counterexample to the idea that band-minimum degeneracy alone guarantees low-density ferromagnetism.

More generally, our findings establish an important constraint for the design of low-density ferromagnetic states in lattice systems. Engineering degenerate band minima is not, by itself, sufficient. A sufficiently strong band-bottom density-of-states singularity is an indispensable condition for stabilizing low-density ferromagnetism in the thermodynamic limit.

The numerical data necessary to reproduce all figures of the present work are available online \cite{DatasetZenodo}.

\begin{acknowledgments}
We acknowledge the use of computational resources from the parallel computing cluster of the Physics and Astronomy Department in University of Bologna. J.W. is supported by the China Scholarship Council(CSC) under Grant No. 202507820051.
\end{acknowledgments}

\bibliography{apssamp}

\end{document}